# Towards a Better Understanding of Online Influence: Differences in Twitter Communication Between Companies and Influencers

## Completed research paper

**Diana C. Hernandez-Bocanegra**
RTG User-Centred Social Media
University of Duisburg-Essen
Duisburg, Germany
Email: diana.hernandez-bocanegra@uni-due.de

**Angela Borchert**
RTG User-Centred Social Media
University of Duisburg-Essen
Duisburg, Germany
Email: angela.borchert@uni-due.de

**Felix Brünker**
RTG User-Centred Social Media
University of Duisburg-Essen
Duisburg, Germany
Email: felix.bruenker@uni-due.de

**Gautam Kishore Shahi**
RTG User-Centred Social Media
University of Duisburg-Essen
Duisburg, Germany
Email: gautam.shahi@uni-due.de

**Björn Ross**
School of Informatics
University of Edinburgh
Edinburgh, United Kingdom
Email: b.ross@ed.ac.uk

## Abstract

In the last decade, Social Media platforms such as Twitter have gained importance in the various marketing strategies of companies. This work aims to examine the presence of influential content on a textual level, by investigating characteristics of tweets in the context of social impact theory, and its dimension immediacy. To this end, we analysed influential Twitter communication data during Black Friday 2018 with methods from social media analytics such as sentiment analysis and degree centrality. Results show significant differences in communication style between companies and influencers. Companies published longer textual content and created more tweets with a positive sentiment and more first-person pronouns than influencers. These findings shall serve as a basis for a future experimental study to examine the impact of text presence on consumer cognition and the willingness to purchase.

# 1    Introduction

Influencer marketing is a marketing strategy often used in recent years on social media. The total amount spent in 2019 is estimated to be $8 billion (Schomer 2019). In addition to traditional advertising, influencer marketing is used by companies to increase their organic reach. Influencer marketing is the process in which companies pay individuals who can reach a large audience to promote their products in an authentic way (Carpenter Childers et al. 2018). Such individuals are known as influencers (Woods 2016). They either work independently or are employed by influencer agencies. Influencer agencies connect influencers and companies not only for regular product advertising but also for commercial events such as Black Friday[1].

Many studies have examined the impact of influential users on consumer purchases, for example through electronic word of mouth (Lin and Wang 2018; Li and Wu 2018), and how to identify influential accounts on social media as possible opinion leaders. However, this is different from the concept of professional influencers, which has evolved into a distinct form of advertisement alongside sponsored posts. Professional influencers are paid by companies to promote a certain product or brand. It is still unclear how these influencers differ from traditional marketing in the way they communicate on social media. Since they are especially hired because of their perceived authenticity, more research into their communication style is necessary. How is their communication characterized and how does it differ from the communication of companies?

To examine the differences in the communication of influencers and brands, we consider social impact theory. According to Latané (1981), influence depends on the dimensions of *strength,* which describes characteristics of an influence source, *immediacy,* meaning the closeness of the source, and the *number* of sources. Thus, considering the dimensions of social impact theory might allow us to reveal meaningful insights about online influencers, influencer marketing, brands and traditional social media marketing practices by brands. Such findings are relevant to better understand their influence on consumer perception or the willingness to purchase an advertised object. This is especially crucial to know for social media managers or influencers, who are interested in optimising their social media strategies.

In order to examine the communication of influencers and companies in a structured way, we tackle one dimension of the social impact theory at a time. In this work, the focus is on the dimension of immediacy. Immediacy is not only defined as the temporal and geographical proximity of a source (such as an influencer, a company).     It can also be conveyed textually in a social media post (Miller and Brunner 2008). Immediacy in the sense of textual presence can reflect physical or psychological presence. Proximity can be expressed, for example, by a high amount of words or large text blocks in social media content (Miller and Brunner 2008). Immediacy as a psychological construct can also be related to the use of paralanguage (e.g. emojis) or sentimental language (Poole, 2000; Rourke et al. 1999). Investigating the dimension of immediacy will broaden the understanding of the textual     presence of social media users during a commercial event, by revealing distinct characteristics of content generated by influencers and companies. For that reason, we consider Twitter content by the most influential accounts – in other words, by those accounts whose tweets were shared the most online and which thus reached many people. We analyse underlying text features of such content in order to draw conclusions about differences in author immediacy between influencers and companies. This addresses the following research question:

**RQ:** How does textual content generated by influential companies and influencers differ on Twitter during a commercial event?

As an analysis approach, we have focused on tweets that were published during Black Friday 2018. We assume that especially commercial events such as Black Friday are an opportunity for companies to show an online presence and make use of influencer marketing to boost their social impact.

# 2    Background

## 2.1    Social Impact Theory

The social impact theory defines influence as "any of the great variety of changes in physiological states and subjective feelings, motives and emotions, cognitions and beliefs, values and behaviour, that occur in an individual, human or animal, as a result of the real, implied, or imagined presence or actions of other individuals" (Latané 1981, p. 343). Moreover, the social impact theory distinguishes influence in the three dimensions; *strength*, *immediacy*, and *number*. While strength focuses on the characteristics of a source of influence, immediacy describes time and space proximity to the target of influence. The number dimension denotes the quantity of influential sources (Latané 1981).

---

[1] https://www.tribegroup.co/blog/black-friday-looks-bright-with-influencer-marketing?

The presence and actions of companies and influencers on social media are assumed to lead to changes in attitude or behaviour of their recipients, that is, social media users (Woods 2016). Regarding the dimension of strength, assertiveness and exaggeration are identified as characteristics predictive of influential users in the online context (Miller and Brunner 2008). On the other hand, Miller and Brunner (2008) understood immediacy as the presence of published content. The presence of textual contributions can be characterised by their total number and the number of words contained in each contribution. The former was found to be a predictor for influence in anonymous, synchronous and collaborative online communication. We assume that such immediacy features, as well as other dimensions of the social impact theory, have an effect in a commercial context, too. Therefore, we focus on immediacy in this work in progress (RQ) by taking a closer look at features connected to text presence, for example, length of the contribution, and the use of frequent words or elements such as emojis.

## 2.2   Influence on Social Media

Since social media have become a stage for everyday life, individuals who exert a strong influence on others have attracted the attention of companies for marketing purposes as well as the attention of researchers. One term, which long predates social media and goes back to the idea of opinion leaders from theories on public opinion formation such as Lazarsfeld's two-step flow model, is that of "influentials" (Watts and Dodds 2007). Similarly, research has termed individuals "influencers" who have a high connectivity to others or who are in a central position in the (social) network that allows them to catalyse a cascade of influence (Bakshy 2011). Kempe et al. (2003) still envisaged these "influential" users as promoting a product for free by recommending it to their friends, after being given a free sample and liking it.

However, the contemporary concept of an influencer, as in influencer marketing, is slightly different. The influencers advertising products on Twitter or Instagram are not simply telling their peers that they enjoyed a particular product, but they are paid to do so, yet they still strive to appear relatable and authentic (Newlands and Lutz 2017). Their terms are negotiated with the advertising agency and specified in a written contract. The amount of money influencers make can be high – as much as $20,000 from a single contract in some instances (Carpenter Childers et al. 2018). As a result, "influencer" is no longer only a label applied by academics to accounts with a high follower count, but it is considered a distinct profession. There are social media users who aspire to become professional influencers, and books (Graham, 2019; Welch, 2019) and events (Stoldt et al. 2019) that promise to show them how to do so. We focus on the influential communication of these individuals who have a high connectivity to other users and who are paid by companies to promote a product, who can also be described as "professional influencers".

Hashtags, mentions, URLs, emojis and the number of words are among the elements that are often used to examine online communication, and that have an effect on the influence of content (Cossu et al. 2015). Miller and Brunner (2008) found that a higher number of words within a text contribution corresponds to social influence in anonymous, collaborative online communication. Findings show that URLs whose webpages evoke positive feelings or which are assessed as interesting are more likely to spread, although a prediction of influence concerning URLs has been found to be unreliable (Bakshy et al. 2011). Hashtags symbolise social influence expressed by the so-called *neighbourhood effect*. It describes that if individuals within a network show the adoption of a trend by using content-related hashtags, others tend to be influenced in their behaviour and also join the community (Backstrom et al. 2006). Mentioning other users (mentions) has been considered to have an impact on sentimental influence in social media (Wu and Ren 2011). Moreover, findings regarding emojis have shown that they promote perceived playfulness in text messages, which positively affects electronic word-of-mouth and is highly related to online influence (Hsieh and Tseng 2017).

Another approach for examining textual contributions on social network sites is analysing communication styles. Linked to personality traits, communication styles may present the author's stable individual predispositions during communication (Page et al. 2013). Research results on the relationship between personality and influence have been somewhat conflicting (Winter et al. 2020).

However, the use of specific words such as pronouns (e.g. I, me, we, they) has been related to personality, social skills, leadership ability and the quality of relationships between people (Pennebaker 2011). Quercia et al. (2011) considered categories that reflect language use for examining influential tweets. High use of first-person pronouns in communication correlates with the personality trait of neuroticism (Stirman and Pennebaker 2001) and is related to self-focus (Quercia et al. 2011). In comparison, second and third-person pronouns symbolise social engagement (Rude et al. 2004). Moreover, Quercia et al. (2011) conducted a sentiment analysis to test the relationship between expressed sentiment on the level of communication styles and influence on Twitter. They found that influential users tend to express a sense of community and negative emotions by often using second- and third-person pronouns as well as words with a negative connotation.

## 3     Research Design

### 3.1    Case Description

The "Black Friday" is a commercial event on the day after Thanksgiving. Over the years, Black Friday has evolved into one of the most popular shopping events in Western culture. Social media can be deployed to distribute marketing communication to further promote this event, which can be enhanced by influencer marketing. Since these marketing strategies affect purchase behaviour and the willingness to purchase, we consider data regarding the Black Friday as relevant to gain new insights in the commercial context.

### 3.2    Data Collection

This case study examines influential content during a commercial event. To this end, we have performed a sequence of steps. First, the study builds upon Twitter communication related to the commercial phenomenon of "Black Friday" (see also Brünker et al. 2020). The data was gathered with a self-developed Java crawler using the Twitter4J library. The collected tweets contained at least one of the following hashtags: *"BlackFriday", "CyberMonday", "BlackFridayDeals", "BlackFriday18", "BlackFridaySale", "BFCM", "BFCM18", "BlackFriday2018"*. These terms were chosen based on Twitter's trending hashtags and comparison to past Black Friday related Twitter communication. Due to the surrounding event "Cyber Monday Week", the hashtags *"CyberMonday2018"* and *"CyberMonday18"* were also considered as relevant indicators for commercial communication during the period. Furthermore, only tweets declared by Twitter as English were considered. In total, we gathered Twitter communication on the Black Friday, from 23 November 2018 00:00 to 23 November 2018 23:59. The selection of keywords was based on the usage frequency during the event, as well as hashtags appearing in the Trends section on Twitter. The tracking yielded a total of 392,606 users creating 542,551 tweets.

In order to identify influential users and their content, we used social network analysis to compute the degree centrality of each user, particularly the in-degree. We defined influential users to be those users who were retweeted the most within the dataset (Oh et al. 2015). In order to cover the majority of shared tweets, we considered the 200 most retweeted users during the examined time period, with a resulting number of 808 tweets for further analysis. These highly retweeted users are responsible for the majority of online communication in the examined case – one user may be retweeted thousands of times and is therefore represented several times in the dataset (Stieglitz et al. 2017). The reason we considered the top 200 retweeted users is the distribution of retweets in the dataset. A few users were responsible for the majority of retweets; thus, to catch the majority of influential users, we focused on the most retweeted ones and excluded the cases belonging to the long tail from the analysis.

To facilitate the analysis of data, we first manually classified users into companies, influencers and other specific roles. Similar suitable categorisations such as influential individual, media or promotional account have been used before to precisely analyse influencers (Bokunewicz & Shulman 2017). Therefore, we followed an inductive category formation (Mayring 2014), and manually checked each of the Twitter profiles. This check involved a number of aspects of the profile, for example the content of profile description, URLs linked as personal websites or connected social media platforms (e.g. Instagram, YouTube, or blogs) as well as a sample of the last tweets shared. We started with one category and benchmarked each account against the criteria of the category. Following that, we either classified the account into the existing category or created a new one. For example, we assigned users to the category of *influencers* who have a high connectivity to other users and are likely to be paid by companies to promote a product. Figure 1 shows two examples. This step involved two independent researchers who came to a substantial agreement of Cohen's kappa = 0.724, signalling a sufficient level of inter-coder reliability (Landis and Koch, 1977).

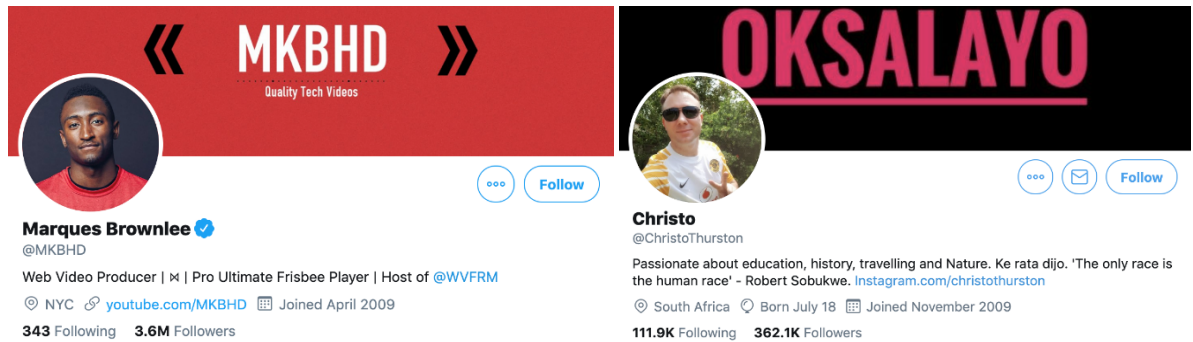

*Figure 1. Examples of Twitter Influencers*

Tweets were preprocessed by removing punctuation, stop words and the hashtags related to Black Friday. In order to identify the differences in communication style between the roles of interest (influencers and companies), we performed a content analysis consisting of the following tasks: determining metrics related to common tweet elements (hashtags, URLs, emojis and mentions), detecting the most frequent words in tweets, and examining communication style (Quercia et al. 2011), especially first-, second- and third-person pronouns. Metrics based on the number of elements such as hashtags, URLs, emojis, and mentions were extracted using regular expressions, character patterns that represent those elements. We use the Python library wordcloud (Mueller 2019) to identify relevant words based on their weighted frequency in the dataset. We also analysed the URLs in the tweets to understand what kind of websites are referenced by companies and influencers with the library pydomains (Sood 2018), which identifies the type of website from a set of predefined categories, such as shopping (e.g. amazon.co.uk), bank, phishing, malware, press, adult, and others.

To further identify communication styles, we extracted the number of pronouns in the first person (e.g. I, we, me, my, myself), the second person (e.g. you, yourself), and the third person (e.g. he, they, his, her). Since pronouns are sometimes omitted in informal language (for example, "just got home" instead of "I just got home"), two coders identified instances of this phenomenon (called ellipsis) in the data set and discussed cases of initial disagreement until they agreed. The reliability of this approach was calculated as κ = 0.66 (first person pronouns), κ = 1.00 (second person), κ = 0.67 (third person) after two coders used the same procedure on a smaller subset (100 tweets). This allows us to calculate the number of sentences that are in the first, second or third person even when the pronoun is absent, and it gives us a measure of how informal the linguistic register used in the tweets is.

In order to compare the content differences between influencers and companies, we performed statistical tests on distributions of length, the number of words, URLs, hashtags, mentions, emojis, and the number of first, second- and third-person pronouns per tweet.

Lastly, we performed a sentiment analysis (cf. Quercia et al. 2011) to classify the sentiment expressed in a text as positive, neutral or negative. The objective is to identify the attitudes and emotions that characterise the content of each tweet. For that reason, we used the tool VADER (Hutto and Gilbert 2014).

## 4    Results

As a first step, we categorised the 200 most retweeted Twitter accounts (authors) into different roles, in accordance with the above-mentioned procedure. After the removal of tweets generated by accounts that were deleted or suspended and, thus, could not be classified at the time of analysis, the total number of tweets that we used for the analysis is 763, by 187 authors. Table 1 shows the number of tweets and authors for each role.

| Role | Examples | # of tweets | # of authors |
|------|----------|-------------|--------------|
| Companies | Cafes and restaurants, airlines, glasses vendors, shops | 264 | 66 |
| Influencers | Professional bloggers, Twitterers, YouTubers, Instagrammers | 54 | 20 |
| Others | Artists, government and politicians, media, individuals, social activists, communities, sports teams | 445 | 101 |
| **Total** | | **763** | **187** |

*Table 1. Categorisation and number of tweets and authors per role*

We aimed to test differences between the two roles regarding length of tweets, and number of elements included within tweets (emojis, hashtags, URLs, and mentions), as well as the use of first, second and third person pronouns. In this respect, and given that data did not fulfill the assumption of normality, we used a nonparametric test for comparison, i.e. Mann-Whitney U test. In addition, chi-square tests were used to compare percentages of tweets with positive, negative and neutral opinions, within each role, as well as the percentages of use of different personal pronouns and the percentages of URL types referenced in tweets within each role. Results indicate that companies tend to write longer tweets, with more hashtags, and emojis than influencers. In fact, tests indicate that:

- The length of tweets (in characters) was significantly greater for companies (Mdn = 154) than for influencers (Mdn = 117), U = 5018, p < .001, d = 0.50, as well as the number of words per tweet, which was significantly greater for companies (Mdn = 24) than for influencers (Mdn = 17), U = 4857, p < .001, d = 0.52. These d values indicate a medium effect size (Cohen 1988). See also Figure 2.1.

- The number of other elements per tweet was also significantly larger for companies than for influencers. This is the case for the number of URLs per tweet (companies: Mdn = 1, 75th percentile = 2; influencers: Mdn = 1, 75p = 1), U = 5558, p = .005, d = 0.01, hashtags (companies: Mdn = 1, influencers: Mdn = 0), U = 5676, p = .012, d = 0.28 and emojis (companies: Mdn = 0, 75th percentile = 1; influencers: Mdn = 0, 75p = 0), U = 5771, p = .006, d = 0.37. Despite the significant difference in the case of URLs, the very small d value indicates a small effect size, therefore the difference can be considered trivial. The d values for emojis and hashtags indicate a small to medium-sized effect. See also Figure 2.2.

- The percentages of tweets with positive, neutral or negative opinions differ significantly between companies and influencers, $X^2$ (2, N = 318) = 9.54, p <.01. See also Figure 2.4.

The analysis of frequent words in the dataset reveals that both companies and influencers frequently communicate about special "deals", "giveaways", and "competitions" where individuals might "win" certain prizes. However, influencers also try to convey messages involving emotions and feelings, as with the frequently used expression "Friday feeling", or messages with a critical intent, such as those related to politics, as evidenced by frequent mentions of "Trump".

The number of mentions per tweet was not significantly greater for companies than for influencers, nor the number of second and third-person pronouns (Mdn = 0, both companies and influencers). However, the results indicate that companies use more first-person pronouns than influencers. The number of first-person pronouns per tweet was significantly greater for companies than for influencers (companies: Mdn = 0, 75th percentile = 1; influencers: Mdn = 0, 75p = 0), U = 5386, p = .002, d = 0.23 (indicating a small to medium-sized effect). Additionally, in the case of companies, the use of first-person pronouns surpasses the use of second and third person pronouns, as depicted in Figure 2.3. This difference was also confirmed using a chi-square test ($X^2$ (2, N = 318) = 17.69, p < .001), which indicates that the distribution of use of pronouns by companies is significantly different from a distribution that assumes that all pronoun types are equally likely to be used. In the case of influencers, this difference is not significant. Moreover, the number of ellipsis (omissions of, for example, pronouns) was higher for companies than for influencers (companies: Mdn = 0, 75th percentile = 1; influencers: Mdn = 0, 75p = 0), U = 5581, p = .004, d = 0.46 (a small to medium-sized effect). We found a positive correlation between the number of ellipsis used and the length of the tweet, r=0.15, p<0.05. However, this correlation can be considered as very weak (Evans 1996).

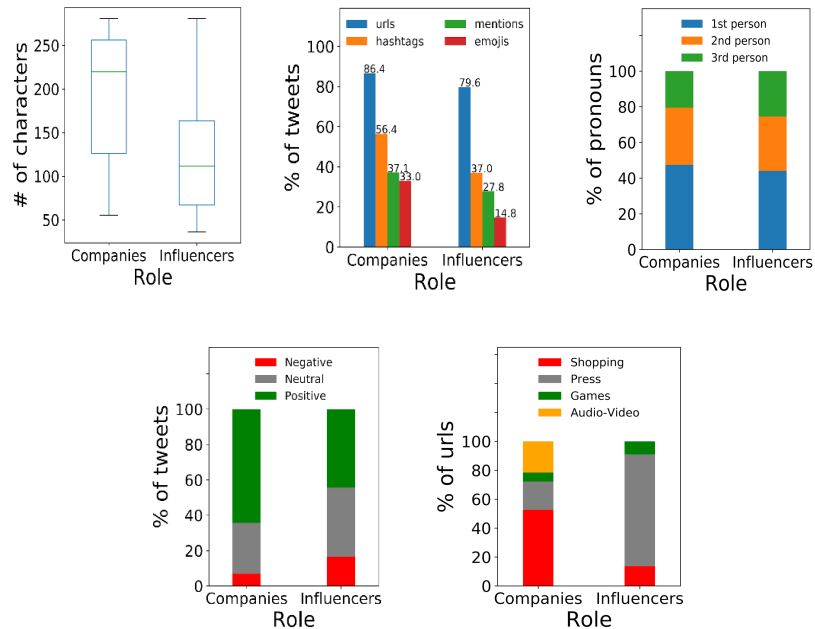

*Figure 2. From left to right, top to bottom: (1) Length of tweets, (2) Elements in tweets, (3) Communicative style (Percentage of each personal pronoun used in tweets), (4) Sentiment analysis (Percentage of tweets with each type of polarity), (5) Types of URLs referenced in tweets.*

The sentiment analysis results show that there is a difference of the proportion of tweets that contain positive, neutral or negative words, between companies and influencers ($X^2$ (2, N = 318) = 9.54, p <.01), as depicted in Figure 2.4. In particular, we found that in the case of companies, the proportion of tweets' polarity is significantly different from a distribution that assumes that the content is tweeted in the same polarity proportion, i.e. 33% of tweets would have each polarity ($X^2$ (2, N = 318) = 73.17, p <.001). In the case of influencers, such difference is not significant.

Lastly, the URL analysis showed that there is a difference in the proportion of URL types referenced in tweets between companies and influencers ($X2$ (4, N = 409) = 33.31, p <.001), as depicted in Figure 2.5. In particular, for influencers, only 3 out of 54 (5.6%) of tweets contained direct links to shopping websites, compared with 51 out of 264 (19.3%) for companies. This difference is statistically significant ($X2$ (1, N= 318) = 6.02, p = .014)). In contrast, of the tweets by influencers, 31.5% contained a link to a press website, compared with 7.2% of tweets by companies ($X2$ (1, N = 318) = 26.33, p< .001). In addition, all 21 links to audio or video hosting websites were posted by company accounts.

## 5    Discussion

The present study exploratively approached the utilisation of Twitter by companies and influencers as well as their communication styles, in the context of a large-scale commercial event. Although the main objectives of influencers and companies are identical, that is, to advertise special deals and competitions, their strategies differ markedly. Examining the findings under social impact theory and its dimension of immediacy, we can deduce that there are indeed differences between companies and influencers regarding characteristics of the text, such as its length or sentiment (we observed that influencers tend to communicate in a more negative way than companies do). The results show that companies use significantly more characters in their tweets, and use emojis and hashtags much more frequently than influencers. In consequence, one can argue that companies are more interested in arousing positive emotions and thus promoting themselves and their products. Therefore, their strategies might be perceived as less authentic (they only highlight positive aspects of their products or services, with no apparent capacity for self-criticism), compared to influencers, who might be perceived as more authentic and credible, since they are more critical in their discussion about products and deals. In particular, most of the negative statements expressed by influencers in the data set involves criticism to unnecessary purchases, e.g. "The trouble with #BlackFriday is you always buy something you don't need just because it's cheap" (sic.). This perceived authenticity has been identified as one of the key characteristics of influencer marketing (Newlands and Lutz 2017). Finally, although the difference in the number of URLs between companies and influencers is negligible, companies are significantly more likely to include direct links to shopping websites and to audio or video hosting websites in their tweets, whereas influencers are more likely to refer to press websites. The latter may be perceived as more neutral and as a more indirect form of promoting a product, further contributing to perceived authenticity.

Considering that another key feature of influencer marketing is that they suggest a personal, almost intimate relationship with their audience (Newlands and Lutz 2017), it is surprising that influencers were less likely to use first-person pronouns in their posts than companies. One possible explanation is that businesses deliberately use many such pronouns in an effort to seem more personable.

Our contributions are novel. Similar research on characteristics of influential tweets did not study them with a focus on the new phenomenon of influencer marketing. For example, focusing on online influence based on a textual level, Miller and Brunner (2008) found that the length of messages positively correlates with social influence. However, their research was conducted on anonymous, collaborative social networks. Context and nature of communication are thus different from those in public social media communication. Therefore, the relationship between online influence and textual characteristics needs to be checked for this domain, as well.

Moreover, Quercia et al. (2011) considered communication styles within influential tweets in a similar manner as we did. However, their approach can rather be seen as connecting the dimensions of immediacy and strength by relating communication styles to specific user types. Instead, we are not addressing such a connection, nor identifying the actual personality of influencers and companies based on their used communication style, but aiming to gain a better understanding of the differences between traditional social media marketing and influencer marketing.

In order to address the effect that communication styles of companies and influencers have on users' perception, we need to look beyond the dimension of immediacy and instead relate to the dimension of strength from social impact theory. It is also conceivable that the impact of used communication styles will multiply the more often influencers or companies publish content to advertise a product. This assumption considers the relation that an interplay of immediacy and number may have to online influence. In summary, a more comprehensive analysis of online influence can be accomplished by considering the effects that the social impact dimensions have on each other.

## 6      Conclusion and Further Steps

As a conclusion, the results suggest various communication strategies as a possible explanation for the identified differences between the two roles. Consequently, in order to analyse the potential impact of such differences on individuals, we suggest examining the perception of content that fits the communication patterns of companies and influential people first.

We plan to expand this research in progress, based on the findings presented in this study. We provide first insights into differences in the communication of influencers and companies. These findings will be used as a foundation for further investigating differences in influential content on Twitter during a commercial event. As future work, we plan to conduct a user experiment, design the test conditions and extract tweets to be compared based on the identified differences between companies and influencers. In this way, we plan to examine how influential content of influencers and companies on Twitter affects the willingness to purchase a product, in order to gain a better understanding about the impact of the dimension of immediacy from social impact theory, in the context of commercial social media communication. Further research in this sense might show the effectiveness of distinct communication strategies during a large-scale commercial event on Twitter.

Based on our preliminary findings, we also plan to examine the dimension of *strength* from social impact theory by investigating the characteristics of companies and influencers and how they are perceived by their recipients. Moreover, the dimension of *number* shall be addressed by considering how the amount of content published by influencers and companies affects the dimensions of strength and immediacy, as well as the willingness to buy an advertised product. In addition, we also plan to extend our analysis to non-textual content. In doing so, we aim to gain a broader view of influence on Twitter during a commercial event under the scope of the social impact theory. Furthermore, it should be noted that our preliminary analysis is limited to the top 200 users using the English language. Future examination may consider original content by more users, as well as cultural differences in the way influential tweets are in other languages. However, the purpose of this study was to analyse the content of the most influential accounts.

## Acknowledgements

This work was supported by the Deutsche Forschungsgemeinschaft (DFG) under grant No. GRK 2167, Research Training Group "User-Centred Social Media".

## Copyright